\documentclass[english]{dfnsikon}
\usepackage[ngerman,english]{babel}
\usepackage{multirow}
\usepackage{todonotes}
\usepackage{subcaption}
\usepackage[utf8]{inputenc}
\usepackage[printonlyused]{acronym}
\usepackage{enumitem}
\usepackage{pifont}
\usepackage[parfill]{parskip}
\usepackage{amsmath}
\usepackage{pgfplots}
\usepackage{multirow}
\usepackage{pdflscape}
\usepackage{caption}
\usepackage{listings}
\usepackage{url}
\usepackage[verbose]{placeins}
\usepackage{breakurl}
\usepackage[breaklinks]{hyperref}

\addto{\captionsngerman}{%

}

\begin{document}

\konferenznummer{31}
\beitrag{A}
\kurzbeschreibung{Geisler et al.: QR Code Phishing Study}

\title{Hooked: A Real-World Study on QR Code Phishing}
\author{Marvin Geisler, Daniela Pöhn, and Wolfgang Hommel\\
       \url{[firstname.lastname]@unibw.de}}
\date{}
\maketitle

\begin{abstract}
The usage of quick response (QR) codes was limited in the pre-era of the COVID-19 pandemic. Due to the widespread and frequent application since then, this opened up an attractive phishing opportunity for malicious actors. They trick users into scanning the codes and redirecting them to malicious websites. In order to explore whether phishing with QR codes is another successful attack vector, we conducted a real-world phishing campaign with two different QR code variants at a research campus. The first version was rather plain, whereas the second version was more professionally designed and included the possibility to win a voucher. After the study was completed, a qualitative survey on phishing and QR codes was conducted to verify the results of the phishing campaign. Both, the phishing campaign and the survey, show that a professional design receives more attention. They also illustrate that QR codes are used more frequently by curious users because of their easy functionality. Although the results confirm that technical-savvy users are more aware of the risks, they also underpin the malicious potential for non-technical-savvy users and suggest further work regarding countermeasures.
\end{abstract}

\section{Introduction}
\label{sec:introduction}

\begin{figure}
     \centering
         \centering
         \includegraphics[width=0.8\linewidth]{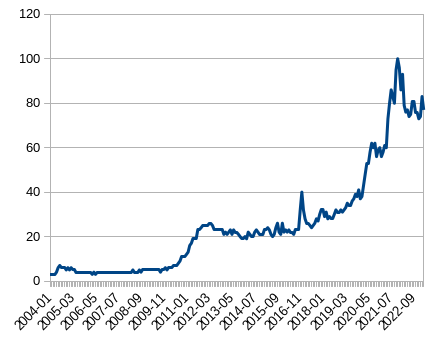}
         \caption{Topic ``QR-Code'' according to Google Trends~\cite{trends}}
         \label{fig:qrcode}
\end{figure}

Quick response (QR) codes are two-dimensional barcodes, which encode different types of information with high density. In 1994, Masahiro Hara invented them to improve the production control of the company Denso Wave~\cite{wahsheh2019secure}. Since then, its usage has become more diverse. For example, QR codes are applied on websites, for marketing, as links for information, or for authentication purposes not only in companies but also in research and education environment. This is the case because they offer simplicity and convenience. Almost all smartphones possess built-in QR code scanners featuring sensors and decoders. During the COVID-19 pandemic, QR codes were frequently applied to, for example, make an appointment for a test, get the COVID-19 test results, an order in a restaurant, or display the vaccination status in the COVID-19 contact tracing app~\cite{9464565}. This increase can also be seen by looking at the trends of search queries by Google Trends~\cite{trends}, as shown in Figure~\ref{fig:qrcode}.

Even though QR codes offer a wide range of advantages, they also pose significant security risks. As QR codes compress the embedded information, they can easily be manipulated or linked to malicious phishing sites~\cite{10.1145/1971519.1971593}. Such malicious QR codes can be printed on flyers or stickers and placed in public places such as universities to lure a wide range of people into scanning them. The quarterly HP Wolf Security Threats Insight Report~\cite{hp} of Q4 2022 warned that QR code phishing has become an almost daily occurrence, especially when used to redirect users to malicious websites asking for credit and debit card details. HP Ireland has not only seen Word documents in Chinese language which claim to entitle users to a government grant, but also phishing campaigns in English masquerading as parcel delivery companies seeking payment. The malicious actors thereby take advantage of generally weaker security protection on smartphones. Even though real-world examples of QR code-based attacks have been reported, only limited research has been conducted -- mostly before the COVID-19 pandemic.

In order to explore if QR code phishing can be launched successfully by attackers and any changes to the studies on QR code phishing before the pandemic can be noticed, we conducted a real-world phishing campaign at the research campus in Garching. We designed two variants of a poster including a QR code and placed them at different frequently visited locations on the research campus. The first version was rather plain, whereas the second poster was more professionally designed and included the possibility to win a voucher. After the study was completed, we conducted a qualitative survey on phishing and QR codes to verify the results of the phishing campaign and get to know, when and why people use QR codes.

Consequently, the contribution of the paper is twofold: 

\begin{enumerate}
\item a comparative QR code phishing study at a research campus; 
\item reasons for and situations when using QR codes based on the study and a survey.
\end{enumerate}

The remainder of the paper is as follows: First, we give an overview of related work in Section~\ref{sec:sota}. In Section~\ref{sec:methodology}, we outline the methodology applied in the QR code study and the survey. The results are summarized in Section~\ref{sec:results}, before we discuss them in Section~\ref{sec:discussion}. Last but not least, we conclude our paper and provide an outlook on future work.

\section{Related Work}
\label{sec:sota}

Franz et al.~\cite{274447} provide an overview of user-oriented phishing interventions and possible future research directions. In their overview of phishing variants, the authors include QR codes as means of accessing information easily. It is noticeable, that comparable less research was conducted on that topic. We first regard influencing factors on the success of phishing attacks, before we focus on the usage of malicious QR codes and specifically QR code phishing.

\subsection{Influencing Factors on Phishing Attacks}

Related to frauds and scams, three taxonomies are often referred to in the literature: Cialdini~\cite{cialdini2007influence} focuses on influence, Gragg~\cite{gragg2003multi} describes triggers, and Stajano et al.~\cite{10.1145/1897852.1897872} analyze the principles of scams. Based on these taxonomies, Ferreira et al.~\cite{10.1007/978-3-319-20376-8_4} provide an overview of different persuasion principles applied in social engineering and specifically phishing based on the analysis of various phishing emails.

In addition, several studies were conducted to reveal the factors influencing the effectiveness of phishing attacks. Downs et al.~\cite{10.1145/1143120.1143131} interviewed 20 people between 18 and 45 years to see whether age had an impact. The authors found no differences. Sheng et al.~\cite{10.1145/1753326.1753383} and Kumaraguru et al.~\cite{10.1145/1572532.1572536} come to contradictory conclusions. However, not all age groups were included in these two studies. In addition, Lin et al.~\cite{10.1145/3336141} included people between 60 and 90 years old in their study. According to the authors, women older than 60 years are the most vulnerable group to phishing. Similar to age, there is no general statement on whether gender influences the success rate. Two studies \cite{10.1145/1753326.1753383,10.1145/1290958.1290968} conclude that women are generally more susceptible to phishing. However, Mohebzada et al.~\cite{6207742} and Diaz et al.~\cite{10.1080/01611194.2019.1623343} found no differences between genders. Several studies~\cite{10.1080/01611194.2019.1623343,10.1145/1290958.1290968} confirm that technical-savvy people are more aware of phishing. In contrast, Alsharnouby et al.~\cite{Alsharnouby2015WhyPS} did not find any connection between technical competence and the identification of a phishing website in their lab study.

\subsection{Malicious QR Codes}

Most studies in the field of phishing attacks concentrate on attacks by email, but there are some studies, which focus on QR codes. Kharraz et al.~\cite{10.1109/DSN.2014.103} and Lerner et al.~\cite{10.1145/2742647.2742650} analyzed QR codes on millions of websites over a period of ten months with the result that the most common form of an attack utilizing a QR code is spoofing on password-protected websites. Kieseberg et al.~\cite{Kieseberg2012} examine the possibility of malicious pixels in QR codes as an attack vector, whereas Zhou et al.~\cite{10.1145/3351284} propose invisible QR code hijacking using smart LED and Dabrowski et al.~\cite{10.1145/2666620.2666624} showcase a barcode-in-barcode attack. Moreover, Krombholz et al.~\cite{10.1007/978-3-319-07620-1_8,10.1109/ARES.2015.84} evaluate possible attacks and challenges for usable security for QR codes as well as countermeasures.

The following studies focus on QR code phishing specifically. Yong et al.~\cite{8843688} provide an overview of current QR code phishing attacks and countermeasures. Mavroeidis and Nicho~\cite{10.1007/978-3-319-65127-9_25} generated a QR code, which led to a fake Gmail website. Based on the results, the authors identified factors that can cause the disclosure of personal data. Vidas et al.~\cite{10.1007/978-3-642-41320-9_4} show that 61\% of the QR codes in the distributed flyers were scanned at least once and 58\% of the participants paid attention to the stored Uniform Resource Locator (URL) of the QR code. Sharevski et al.~\cite{10.1145/3549015.3554172} examine whether QR code phishing can be successful. The authors conducted a study with 173 participants, which could choose between different registration options. 67\% of the participants preferred signing up via Google or Facebook, 18.5\% created a new account, and 14.5\% skipped the process. In an additional survey, the participants argued that convenience is the biggest factor to disclose personal information. Additionally, Kumar et al.~\cite{10.1007/978-3-031-06394-7_64} conducted a survey with 132 people on their awareness, perception, and cyber-hygiene behavior towards QR codes and their malicious usage. While around half of the participants had concerns, they used QR codes nonetheless. More than one third had no concerns at all and were not aware of any security risks associated with QR codes.

\subsection{Summary}

Although some approaches examined various aspects of QR code phishing, these mostly took place before the COVID-19 pandemic and before QR codes were increasingly being used. To control and compare the results from before and now as well as different locations, a new study is required.

\section{Methodology}
\label{sec:methodology}

In the following section, we outline the methodology of our approach, consisting of a QR code study and a user survey related to the usage of QR codes and phishing. After that, we will discuss the limitations of this approach.

\subsection{QR Code Study}

Similar to the methodology of Vidas et al.~\cite{10.1007/978-3-642-41320-9_4}, we placed posters with QR codes in ten public locations, where posters and flyers are routinely placed, on the research campus Garching in Germany. The location was selected because the research campus is frequented by many young people, while students and employees are typically over 18 years old. If permissions were required, we got them beforehand.

\subsubsection{Posters with QR Codes}
The posters suggested that the survey examines the impact of inflation on personal life. Embedded were QR codes that represented URLs to our secured web server with the name \emph{Study-Research (SR)} of our fictitious research institution. To validate that the design of the posters can provide an effect on the success rate, we designed two variants of the posters.

\begin{enumerate}
\item Plain poster, as displayed in Figures~\ref{fig:variant1} and \ref{fig:variant1-real}.
\item Poster with professional design and advertisement, as shown in Figure~\ref{fig:variant2}. This version utilizes the color blue as it offers the greatest trust according to Peters et al.~\cite{peters}. In addition, the text suggested that the participants could win an Amazon voucher, which applied Cialdini's principle of scarceness~\cite{cialdini2007influence}.
\end{enumerate}

We posted them secured in cling films for two consecutive weeks at the beginning of December 2022. The cling film was added due to the weather conditions of posters on the outside and did not restrict the usage of the QR codes. This can be seen in Figure~\ref{fig:variant1-real}. The posters were taken off on the last day.

\subsubsection{Data Acquisition and Ethical Considerations}
\begin{figure*}[!htpb]
     \centering
     \begin{subfigure}[b]{0.45\textwidth}
         \centering
         \includegraphics[width=\textwidth]{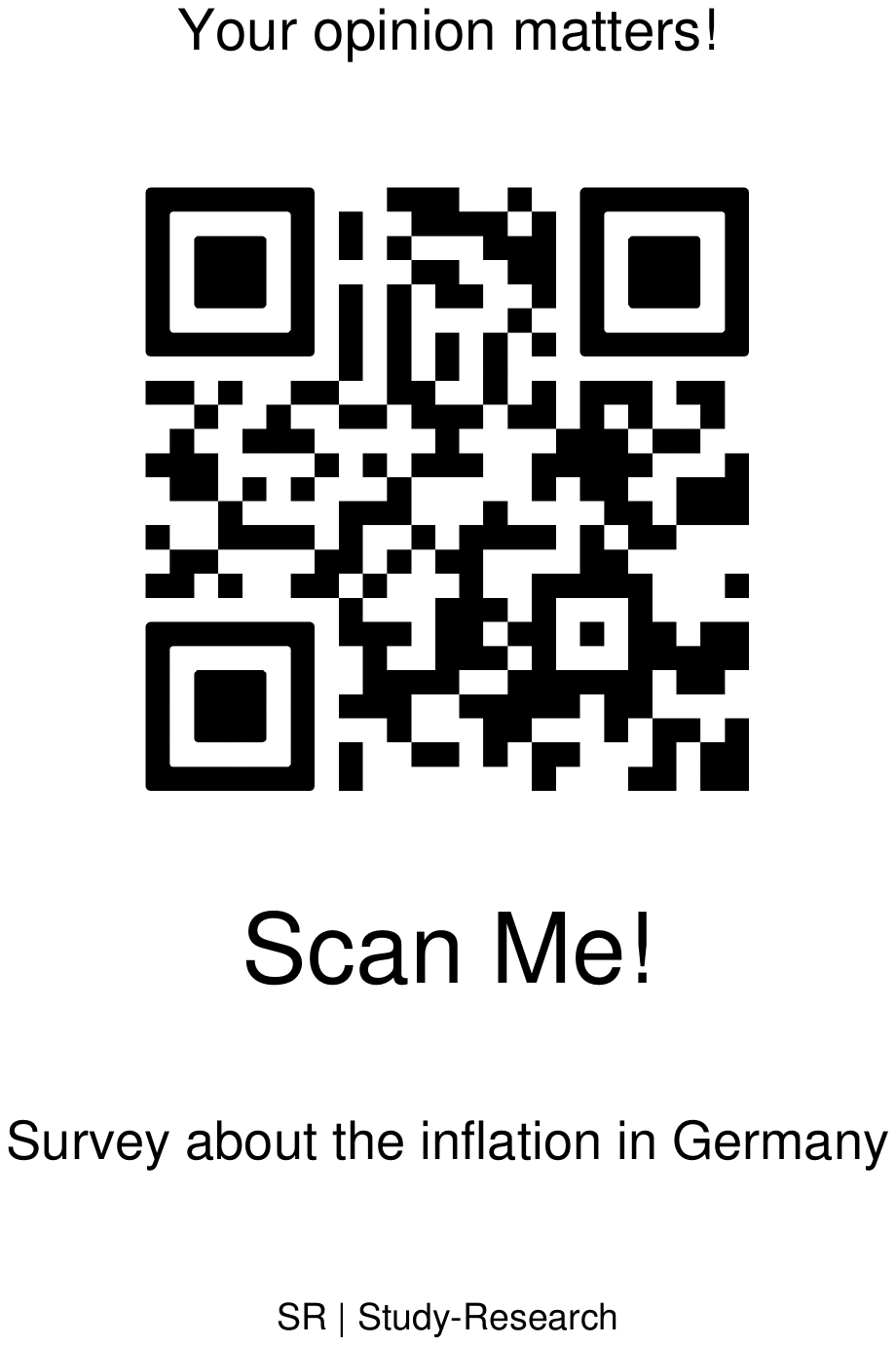}
         \caption{Design of variant 1}
         \label{fig:variant1}
     \end{subfigure}
     \hfill
     \begin{subfigure}[b]{0.45\textwidth}
         \centering
         \includegraphics[width=\textwidth]{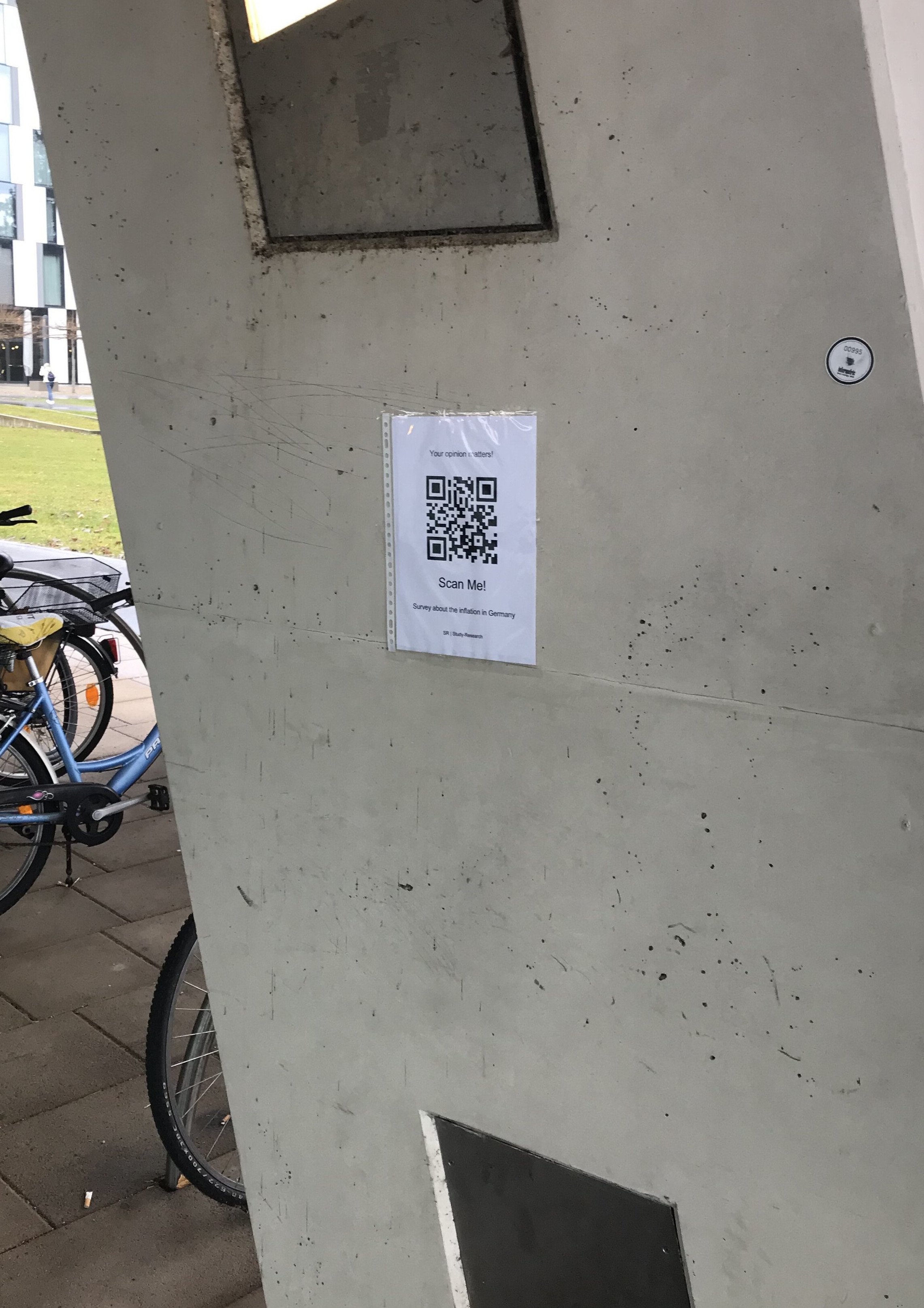}
         \caption{Variant 1 posted on a wall}
         \label{fig:variant1-real}
     \end{subfigure}
\hfill
     \begin{subfigure}[b]{0.45\textwidth}
         \centering
         \includegraphics[width=\textwidth]{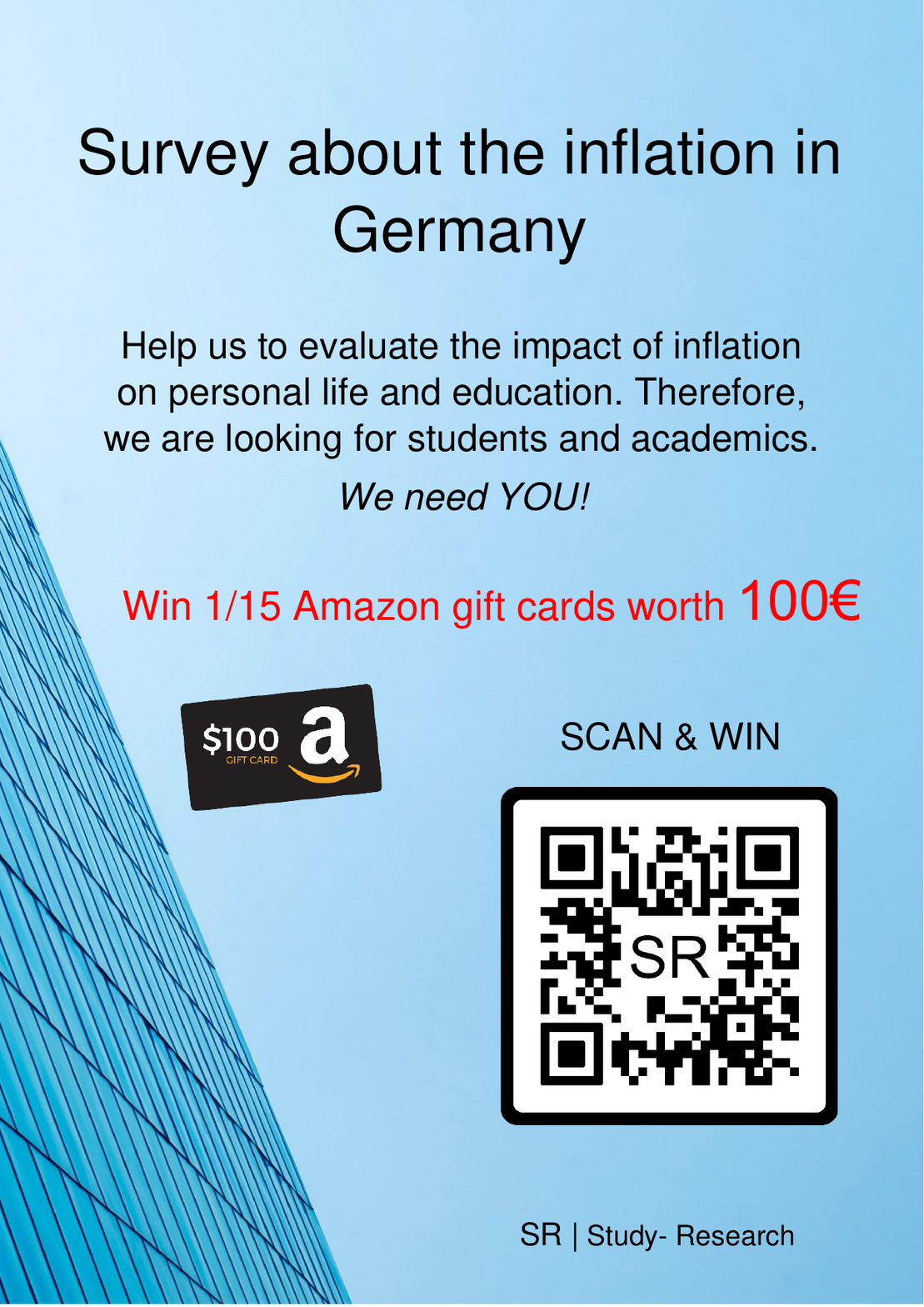}
         \caption{Design of variant 2}
         \label{fig:variant2}
     \end{subfigure}
     \hfill
     \begin{subfigure}[b]{0.45\textwidth}
         \centering
         \includegraphics[width=\textwidth]{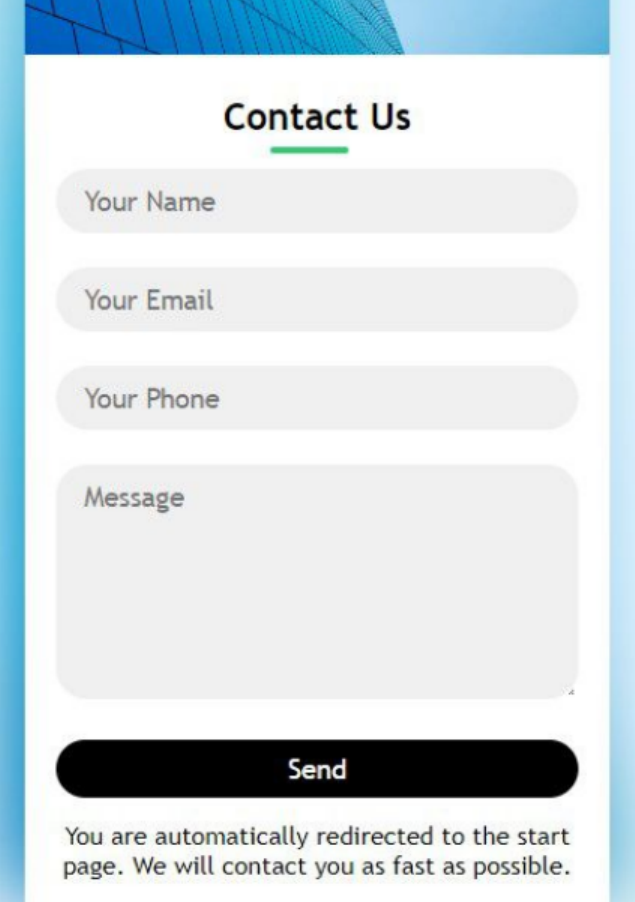}
         \caption{Contact form of the website}
         \label{fig:variant2-contact}
     \end{subfigure}
        \caption{Poster variants}
        \label{fig:variants}
\end{figure*}

If a participant scanned the QR code, they were taken to our website where they could try to register for the survey and were informed about the experiment. The website used the same design as poster variant 2. We recorded the scanning of the QR code, the access time, the click to start the survey, and -- if entered -- a hash of the email address. As the participants scanned the QR code and followed the URL to participate in a study, they were indirectly agreeing, although the different topic was misleading. After the experiment, the participants were shown information about the study. In addition, contact information and a form were accessible, as seen in Figure~\ref{fig:variant2-contact}. The study was approved by the ethical committee of the university.

\subsection{Survey}

In addition, we designed a survey to get an understanding of the reasons for scanning QR codes. The participants were recruited by sharing the link to the survey with an introductory text with friends, colleagues, and further persons. We focused on the same age, occupation, and education as well as the characteristics of technical-savvy as we concluded from the research campus. The online survey was conducted after the QR code study from mid-December 2022 to the beginning of January 2023.

The study consisted of:

\begin{enumerate}
\item Introduction;
\item Consent form;
\item Socio-demographic questions;
\item Preferences of the variants;
\item Reasons for the preferences and usage;
\item Generic questions;
\item Experiences with phishing;
\item Countermeasures against phishing;
\item Closing and note of thanks. 
\end{enumerate}

The data was stored anonymously. After the survey time has ended, we analyzed the data in regard to our research questions. The findings were also compared with the results of the QR code study. The survey was in compliance with the standards of the university review board and, consequently, did not require specific approval.

\section{Results}
\label{sec:results}

In this section, we describe the results of the QR code study (see Section~\ref{sec:r-qr}) and user survey (see Section~\ref{sec:r-user}).

\subsection{QR Code Study}
\label{sec:r-qr}

The results of the QR code study are summarized in Table~\ref{tab:results-study}. Values in ``( )'' represent numbers, which can be traced back to unique participants, as some tried to scan the QR code, start the survey, and enter the email address several times. In the end, 45 times an email address was entered by 30 unique participants, of which five used variant 1 (16.67\%) and 25 variant 2 (83.33\%). Thereby, variant 2 attracted significantly more participants. Compared to the number of potential targeted persons on the research campus with 7,500 employees and 17,500 students~\cite{gate}, the number of actual participants of the QR code phishing study is very small (less than 1\%).

\begin{table}[!htpb]
\caption{Results of the QR code study}
\label{tab:results-study}
\centering
\begin{tabular}{lcccc}
\toprule
\textbf{Variant}    & \textbf{Scans} & \textbf{Survey} & \textbf{Email} & \textbf{Multiple} \\ \midrule
Variant  1 & 13 (12)   & 8 (7)          & 6 (5)         & 1              \\
Variant  2 & 38 (25)   & 39 (25)        & 39 (25)       & 7              \\ \midrule
Sum        & 51 (37)   & 47 (32)        & 45 (30)       & 8            \\ \bottomrule
\end{tabular}
\end{table}

\begin{figure*}[!htpb]
     \centering
     \begin{subfigure}[b]{0.6\textwidth}
         \centering
         \includegraphics[width=\textwidth]{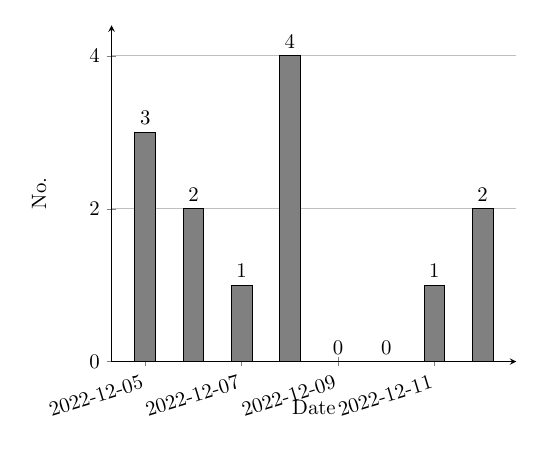}
         \caption{Scans per dates of variant 1}
         \label{fig:dates1}
     \end{subfigure}
     \hfill
     \begin{subfigure}[b]{0.6\textwidth}
         \centering
         \includegraphics[width=\textwidth]{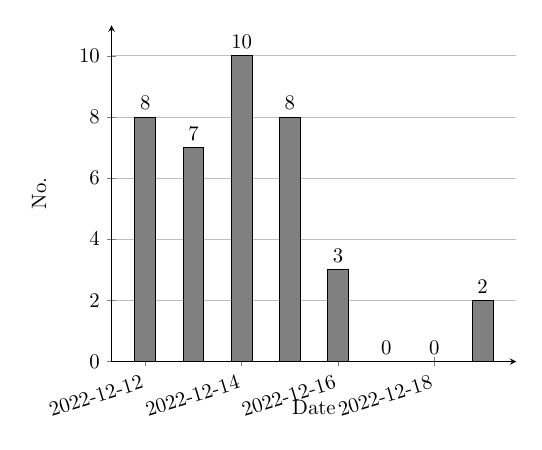}
         \caption{Scans per dates of variant 2}
         \label{fig:dates2}
     \end{subfigure}
        \caption{Number of scans over time}
        \label{fig:dates}
\end{figure*}

It is noticeable that variant 2 was once more often started than scanned, which could be the case by reloading the website without re-scanning the QR code. Figure~\ref{fig:dates} shows the number of scans over time for variant 1 (see Figure~\ref{fig:dates1}) and variant 2 (see Figure~\ref{fig:dates2}). Besides the weekends (days 6 and 7), no clear tendency can be noticed.

\subsection{Survey}
\label{sec:r-user}

After the QR code study, the survey was conducted. 125 participants took part in the survey. The data was stored anonymized. All participants, which received more than 100 malus points (negative points are given, for example, for extremely fast, straight, and weird completion~\cite{sosci}) were excluded as such values are classified as substandard. After data cleansing, we had 123 participants with an average of 30.3 years of age (minimum: 18, maximum: 60), of which 58 (47.16\%) were female, 64 (52.03\%)  male, and one (0.81\%) diverse. 33.33\% were students or trainees, whereas 60.98\% had finished their studies. All participants lived in Germany during the time of the study.

\subsubsection{Preferences of a Variant}

First, we asked about the preferences of a variant and the reasons for it. While, 37 participants (30.08\%) would choose variant 1, 55 participants (44.71\%) would select variant 2. However, 31 participants (25.20\%) said that they would not scan any QR code. The result is shown in Table~\ref{tab:pref}.

\begin{table}[!htpb]
\caption{Preference of a poster variant}
\label{tab:pref}
\centering
\begin{tabular}{lcc}
\toprule
\textbf{Poster} & \textbf{No} & \textbf{Percentage} \\  \midrule
Variant 1         & 37         & 30.08\%     \\
Variant 2             & 55        & 44.71\%     \\
None                & 31         & 25.20\%        \\ \bottomrule
\end{tabular}
\end{table}

Based on these answers, the reasons for scanning one or the other QR code were asked. An overview of the reasons for both variants can be found in Table~\ref{tab:reasons}. For variant 1, trustworthiness and size, and design of the QR code scored the highest with 48.65\% each. The most frequently mentioned reasons for variant 2 were the included voucher (47.27\%) and professionalism (43.64\%). Interestingly, the size and design of the QR code scored high in both variants. 24 participants (77.42\%) named the lack of trust as the main reason for not scanning either variant. In addition, ``not interested'' (25.80\%), ``unresponsive'' (22.58\%), ``safety concerns'' (4.17\%), and insufficient information on sources/author (4.17\%) were named.

\begin{table}[!htpb]
\caption{Reasons for preferring a poster variant}
\label{tab:reasons}
\centering
\begin{tabular}{lcccc}
\toprule
\multirow{2}{*}{\textbf{Reason}} & \multicolumn{2}{c}{\textbf{Variant 1}} & \multicolumn{2}{c}{\textbf{Variant 2}} \\
                        & No.        & Percentage        & No.        & Percentage        \\ \midrule
Professionalism         & 9         & 24.32\%           & 24        & 43.64\%           \\
Trustworthy             & 18        & 48.65\%           & 16        & 29.10\%           \\
Interest                & 4         & 10.81\%           & 6         & 10.91\%           \\
Size and design         & 18        & 48.65\%           & 21        & 38.18\%           \\
Attractiveness          & 4         & 10.81\%           &           &                   \\
Others                  & 1         & 2.7\%             & 4         & 34.54\%           \\
Voucher                 &           &                   & 26        & 47.27\%           \\
Topic                   &           &                   & 19        & 34.54\%          \\ \bottomrule
\end{tabular}
\end{table}

\subsubsection{Interaction with QR Codes}

Another goal of the survey was to find out when, how, and why people interact with QR codes. The participants were asked for the reasons and situations, in which they utilize QR codes. The results are summarized in Table~\ref{tab:location}. 16 participants (13.01\%) stated that they generally do not scan QR codes. However, the functionality and accessibility of a QR code was stated as the main reason (70.73\%) for the use of QR codes by the participants, such as using WiFi at friends via QR code scanning. The second most common reason for using QR codes is curiosity (26.02\%). Regarding the most frequently stated situations in which QR codes are being used, the participants named the use of COVID-19 tests (65.85\%), followed by restaurant visits (63.41\%), university environment (34.15\%), and the use in public facilities, for example, for opening hours (32.52\%). 

\begin{table}[!htpb]
\caption{Reasons and situations for using QR codes}
\label{tab:location}
\centering
\begin{tabular}{lclc}
\toprule
\textbf{Reason} & \textbf{Percentage} & \textbf{Situation} & \textbf{Percentage} \\ \midrule
Functionality   & 70.73            & COVID-19 tests    & 65.85            \\
Curiosity       & 26.02            & Restaurant visit  & 63.41            \\
\textbf{}       &                  & University environment       & 34.15            \\
Not scanning at all                &   13.01               & Public facilities & 32.52           \\ \bottomrule
\end{tabular}
\end{table}

\subsubsection{Experiences with Phishing}

Asked whether they knew beforehand that phishing using QR codes is possible, 50.41\% answered ``yes'' and 47.15\% ``no''. Only three participants chose ``no answer'' (2.44\%).

Next, we wanted to know about their experiences with phishing. 80 people (65.04\%) stated that they had rarely or never been affected by phishing. On the other hand, 24 people (19.51\%) answered that they have often or very often experienced phishing. 19 participants answered the question with ``sometimes'' (19.45\%).  The experienced phishing types are via email (63.41\%), social media (30.01\%), telephone (29.27\%), and short message service (SMS) (26.02\%). 39 participants (31.71\%) had no contact with any of the variants just mentioned. The experiences with the different variants are summarized in Table~\ref{tab:variant}.

\begin{table}[!htpb]
\caption{Experiences with phishing variants}
\label{tab:variant}
\centering
\begin{tabular}{lcc}
\toprule
\textbf{Variant}         & \textbf{No.} & \textbf{Percentage} \\ \midrule
No              & 39 & 31.71\%    \\
Email           & 78 & 63.41\%    \\
Phone (Vishing) & 36 & 29.27\%    \\
SMS (SMiShing)  & 32 & 26.02\%    \\
Social Media   & 37 & 30.01\%    \\
Others          & 0  & -         \\ \bottomrule
\end{tabular}
\end{table}

Last but not least, we asked about countermeasures. 58 participants (47.15\%) knew different measures, whereas 30 participants (24.39\%) answered with no, and 32 participants (26.02\%) were not sure. Following this question, the majority of the participants requested more information about countermeasures.

\section{Discussion}
\label{sec:discussion}

In order to conduct a successful phishing campaign with QR codes, several aspects have to be obeyed, such as the actual design, topic, and text as well as the place to launch the posters. Since phishing is a scam variant, the aspect of trustworthiness must be taken into account during the design of the poster and the website. We incorporated blue colors, the usage of hypertext transfer protocol secure (HTTPS), a contact form, and the terms of service into the website. In addition, the Amazon voucher worked as a hook. QR codes can be used for phishing campaigns, however, several factors have to work together to receive enough victims. Based on our study and survey, we conclude that at least technical-savvy people are less likely to fall for QR code phishing.

\subsection{QR Code Study}

Although the entire QR code study was conducted at the same place, there are differences in the results. It can be seen that the second variant received more attention, resulting in the QR code being scanned more often. This shows that design and the embedded QR code have an impact on user behavior. Overall, the second variant was scanned almost three times as often as the first variant.

As we did not count the number of persons passing by the posters, we cannot conclude how many actually have noticed the QR code and cannot compare this number with the actual scans. Based on the number of students and employees, the number of scans is comparably low. In comparison to Vidas et al.~\cite{10.1007/978-3-642-41320-9_4}, our QR codes were scanned much less. One reason could be that the topic was not interesting enough. Due to the  reason that the study took place at a research campus with technical-savvy persons, it could be that these persons are more aware of possible cybersecurity attacks. This might also be true since QR code phishing is happening more frequently and, hence, might be known by the persons. However, this aspect is in contrast to Kumar et al.~\cite{10.1007/978-3-031-06394-7_64} findings.

We applied the same URL and the same website for all posters. Therefore, we cannot see differences regarding the locations. In addition, participants who already scanned the first QR code might have clicked away when seeing the same website in the second variant. In order to receive more participants, more locations as well as posters could be used. However, this might be ethical questionable as it could include interactions with minors.

Jagatic et al.~\cite{10.1145/1290958.1290968} focused on emails and received a higher success rate. A comparison with emails including QR codes could share lights on the level of awareness. As we did not actually phish due to ethical recommendations and legal considerations, we cannot be certain if the participants would also become QR phishing victims.

\subsection{Survey}

Due to organizational issues, the survey could not be sent to the persons, which might have seen the posters. Similar to Vidas et al.~\cite{10.1007/978-3-642-41320-9_4}, we could have asked those who scanned the QR codes to fill in the survey. However, the survey would not be representative as those who did not scan the QR codes would still be missing. To recruit more participants, the survey was adapted and sent to persons known by the authors. By distributing the survey that way, we tried to adjust the level of technical-savvy persons to the proportion at the research campus. Nonetheless, we cannot say for sure that the participants reacted the same way. Last but not least, the research campus and the survey are based in Germany. The culture in other countries could be different and consequently, studies in different countries would be interesting to compare.

\section{Conclusion and Outlook}
\label{sec:conclusion}

With the frequent usage of QR codes in the COVID-19 pandemic, they have become an integral part of everyday life. Although QR codes are convenient, they also come with security risks. In order to re-evaluate the risks of QR code phishing, we designed and conducted a QR code study at a research campus. Due to the low rate of participants and the results of the survey, we assume that technical-savvy people are comparably more aware of the risks of QR codes. However, only 25.20\% would not scan any QR code and 47.15\% know countermeasures. Hence, scalable awareness campaigns and technical countermeasures are important.

Even though the number of participants was low, we noticed a difference in the design of the posters, the embedded topic, and hooks such as vouchers. Based on the results, we recruited people with the same characteristics for our user study to explore the reasons for using and scanning QR codes. Thereby, we recognized that the voucher can indeed lure people into scanning a QR code. In addition, trustworthiness, size and design of the QR code, and the topic can be reasons to fall for QR code phishing. Hence, variances of phishing, such as QR code phishing, should be included into common awareness training.

In future work, we want to extend our QR code study to include more places, such as more universities, and use specific URLs. Thereby, we hope to get an understanding of the characteristics of a person, which scans those QR codes. In order to facilitate the insights, we want to forward the participants to a user survey. Also, the answers of those who do not scan the QR codes would be interesting. Additionally, we want to compare the results to phishing emails, which include QR codes.

\bibliographystyle{abbrv}
\bibliography{qr-code}

\end{document}